\documentclass{iaus}
\usepackage{graphicx}

\title[The Imprints of IMBHs: Monte-Carlo Simulations] 
{The Imprints of IMBHs on the Structure of Globular Clusters: Monte-Carlo Simulations}

\author[Stefan Umbreit, John M. Fregeau \& Frederic A. Rasio]   
{Stefan Umbreit$^1$, John M. Fregeau$^1$
 \and Frederic A. Rasio$^1$}

\affiliation{$^1$ Northwestern University, Dearborn Observatory\\ 2131 Tech Drive, Evanston, 
Illinois, 60208, USA \\ email: {\tt s-umbreit@northwestern.edu}}

\pubyear{2008}
\volume{246}  
\jname{Dynamical Evolution of Dense Stellar Systems}
\editors{E. Vesperini, M. Giersz, A. Sills eds.}
\begin{document}

\maketitle

\begin{abstract} 

We present the first results of a series of Monte-Carlo simulations
investigating the imprint of a central black hole on the core structure of
a globular cluster. We investigate the three-dimensional and the projected
density profile of the inner regions of idealized as well as more realistic
globular cluster models, taking into account a stellar mass spectrum, stellar
evolution and allowing for a larger, more realistic, number of stars than was
previously possible with direct N-body methods. We compare our results to
other N-body simulations published previously in the
literature.  

\keywords{stellar dynamics, methods: n-body simulations,
globular clusters: general}
\end{abstract}

\firstsection 
\section{Introduction} 

As recently as 10 years ago, it was generally believed that black holes (BHs)
occur in two broad mass ranges: stellar ($M\sim 3-20 M_{\odot}$), which are
produced by the core collapse of massive stars, and supermassive ($M\sim 10^6 -
10^{10} M_{\odot}$), which are believed to have formed in the center of galaxies
at high redshift and grown in mass as the result of galaxy mergers (see e.g.
Volonteri, Haardt \& Madau 2003).   However, the existence of BHs with masses
intermediate between those in the center of galaxies and stellar BHs could not
be established by observations up until recently, although intermediate mass BHs
(IMBHs) were predicted by theory more than 30 years ago; see, e.g.,  Wyller
(1970).  Indirect evidence for IMBHs has accumulated over time from observations
of so-called ultraluminous X-ray sources (ULXs), objects with fluxes that exceed
the angle-averaged flux of a stellar mass BH accreting at the Eddington limit.
An interesting result from observations of ULXs is that many, if not most, of them
are associated with star clusters. It has long been speculated (e.g., Frank \&
Rees 1976) that the centers of globular clusters (GCs) may harbor BHs with
masses $\sim 10^3 \rm{M_{\odot}}$. If so, these BHs affect the distribution
function of the stars, producing velocity and density cusps.  A
recent study by Noyola \& Gebhardt (2006) obtained central surface brightness
profiles for 38 Galactic GCs from HST WFPC2 images.  They showed that half of
the GCs in their sample have slopes for the inner 0.5" surface density
brightness profiles that are inconsistent with simple isothermal cores, which
may be indicative of an IMBH. However, it is challenging to explain the full
range of slopes with current models. While analytical models can only explain
the steepest slopes in their sample, recent N-body models of GCs containing
IMBHs (Baumgardt et al. 2005), might  explain some of the intermediate surface
brightness slopes.

In our study we repeat some of the previous N-body simulations of GCs with central IMBHs
but using the Monte-Carlo (MC) method. This gives us the advantage to model the evolution of
GCs with a larger and thus more realistic number of stars.  
We then compare the obtained surface brightness profiles
with previous results in the literature.

\section{Imprints of IMBHs} \label{ImprintsIMBH}
\begin{figure}
\begin{tabular}{cc} \includegraphics[width=0.45\textwidth,clip]{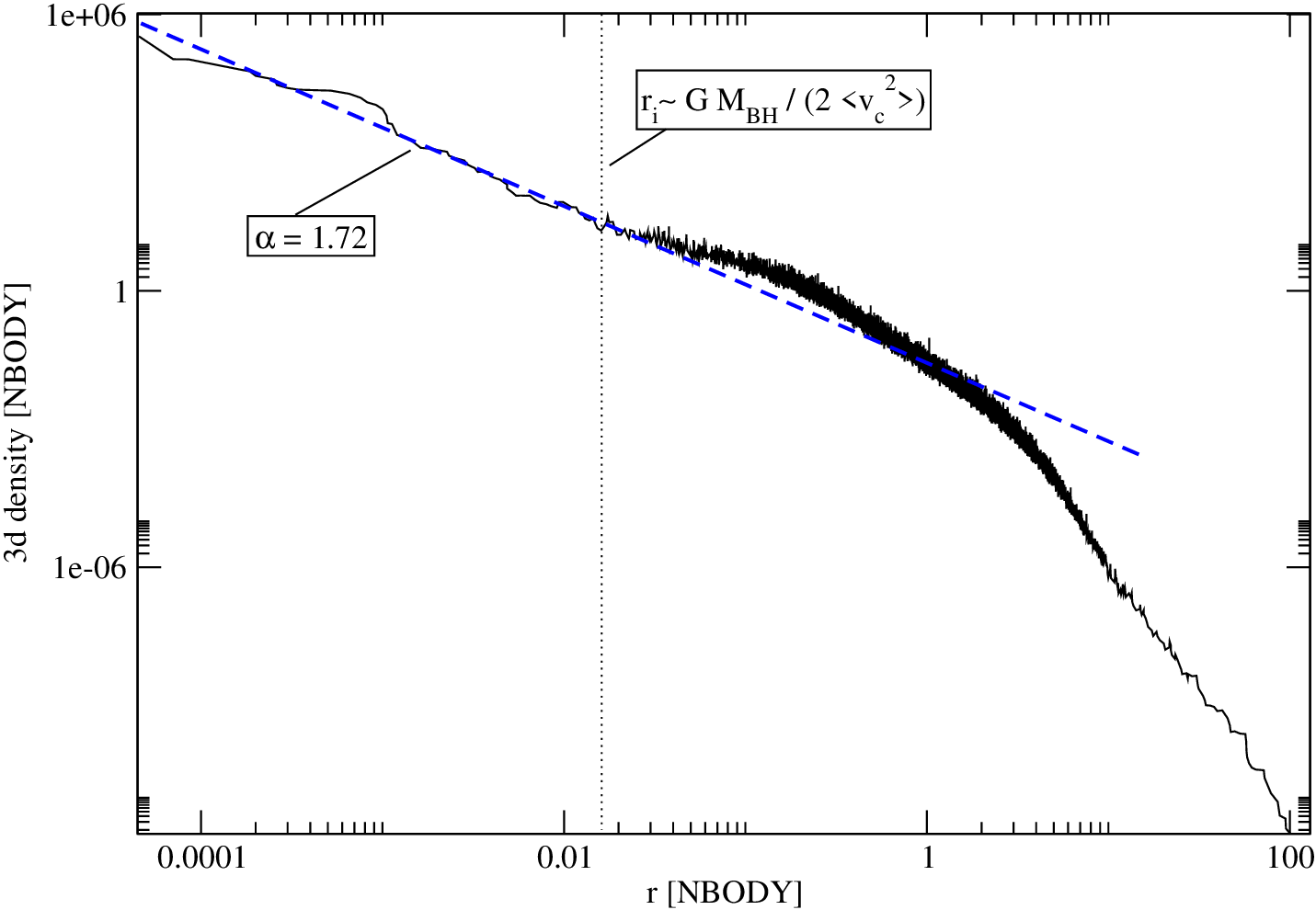}
\includegraphics[width=0.33\textwidth,clip]{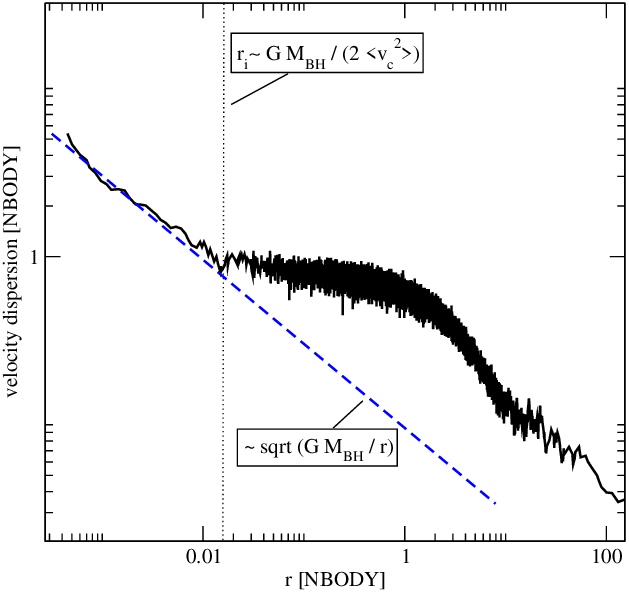}
\end{tabular} 
\caption{The imprint of an IMBH on the stellar distribution of a
GC. On the left-hand side, the radial number density profile of an
evolved single-mass GC is shown (solid) together with a power-law
fit to its inner region (dashed line). The right hand side shows its velocity
dispersion profile (solid) and the Keplerian velocity profile of the IMBH. 
The dotted line marks its radius of influence.} 
\label{Imprints} 
\end{figure}

The dynamical effect of an IMBH on the surrounding
stellar system was first described by Peebles (1972), who argued that the
bound stars in the cusp around the BH must obey a shallow power-law density distribution to account for
stellar consumption near the cluster center. Analyzing the Fokker-Planck
equation in energy space for an isotropic stellar distribution, Bahcall \& Wolf
(1976) obtained a density profile with $n(r)\propto r^{-7/4}$, which is now
commonly referred to as the Bahcall-Wolf cusp. The formation of this cusp has
been confirmed subsequently by many different studies using different techniques 
and also, more recently, by direct $N$-body methods (Baumgardt et al. 2004). 
In Fig. \ref{Imprints} 
such a profile from one of our simulations is shown (for initial cluster
parameters see  Baumgard et al. (2004) (run16)). As can be clearly seen, 
the density profile of the
inner region of the evolved cluster can be very well fitted by a power-law and
the power-law slope $\alpha$ we obtain is, with $\alpha=1.72$, in good agreement
with the value found by Bahcall \& Wolf (1976). Also the extent 
of the cusp profile
is given by the radius where the Keplerian velocity of a star around the central
BH equals the velocity dispersion of the cluster core, the radius of influence of the IMBH. 

However, Baumgardt et al. (2005) found that such a cusp in density might not be
easily detectable in a real star cluster, as it should be much shallower and
difficult to distinguish from a standard King profile. They find that this is
mainly an effect of mass segregation and stellar evolution, where the more
massive dark stellar remnants are concentrated towards the center while the
lower-mass main sequence stars that contribute most of the light are much less
centrally concentrated. In their simulations they found power-law surface
brightness slopes ranging from $\alpha=-0.1$ to $\alpha=-0.3$. Based on these
results they identified 9 candidate clusters from the sample of galactic GCs of
Noyola \& Gebhardt (2006) that might contain IMBHs.
However, the disadvantage of current N-body simulations is that
for realistic cluster models, that take into account stellar evolution and a
realistic mass spectrum, the number of stars is restricted to typically less
than $2\times10^{5}$ as these simulations require a large amount of computing
time.  However, many GCs are known to be very massive, with masses reaching up
to $1\times10^6 \rm{M_\odot}$ resulting in a much larger number of stars one has
to deal with when modelling these objects. In previous N-body simulations,
such large-N clusters have been scaled down to low-N systems.  Scaling down can
be achieved in two ways (e.g. Baumgardt et al.  2005): either the mass of
the central IMBH $M_{BH}$ is kept constant and $N$ is decreased, effectively decreasing
the total cluster mass $M_{Cl}$, or the ratio $M_{BH}/M_{Cl}$ is kept
constant, while lowering both $M_{BH}$ and $M_{Cl}$.  As both $M_{BH}/M_{Cl}$
and the ratio of $M_{BH}$ to stellar mass are important parameters that influence
the structure of a cluster, but cannot be held constant simultaneously when lowering $N$, it
is clear that only with the real $N$ a
fully self-consistent simulation can be achieved. One such method that can
evolve such large-$N$ systems for a sufficiently long time is the
MC method.

\section{Monte-Carlo Method with IMBH}

The MC method shares some important properties with direct $N$-body methods,
which is why it is also regarded as  a randomized $N$-body scheme (see e.g.
Freitag \& Benz 2001). Just as direct $N$-body methods, it relies on a
star-by-star description of the GC, which makes it particularly straightforward
to add additional physical processes such as stellar evolution. Contrary to
direct $N$-body methods, however, the stellar orbits are resolved on a
relaxation time scale $T_{rel}$, which is much larger than the crossing time
$t_{cr}$, the time scale on which direct $N$-body methods resolve those orbits. This
change in orbital resolution is the reason why the MC method is able
to evolve a GC much more efficiently than direct $N$-body methods. It achieves
this efficiency, however, by making several simplifying assumptions: (i) the
cluster potential has spherical symmetry (ii) the cluster is in dynamical
equilibrium at all times (iii) the evolution is driven by diffusive 2-body
relaxation.  The specific implementation we use for our study  is the
MC code initially developed by Joshi et al. (2000) and further
enhanced and improved by Fregau et al. (2003) and Fregau \& Rasio (2007). The
code is based on H\'enon's algorithm for solving the Fokker-Planck equation. It
incorporates treatments of mass spectra, stellar evolution, primordial
binaries, and the influence of a galactic tidal field.

The effect of an IMBH on the stellar distribution is implemented in a manner similar to
Freitag~\&~Benz~(2002).  In this method the IMBH is treated as a fixed, central
point mass while 
stars are tidally disrupted and accreted
onto the IMBH whenever their periastron distances lie within the tidal radius,
$R_{disr}$, of the IMBH. For a given star-IMBH
distance, the velocity vectors that lead to such orbits form a so called
loss-cone and stars are removed from the system and their masses are added to the BH as soon
as their velocity vectors enter this region.  However, as the star's removal
happens on an orbital time-scale one would need to use time-steps as short as
the orbital period of the star in order to treat the loss-cone effects in the
most accurate fashion. This would, however, slow down the whole calculation
considerably.  Instead, during one MC time-step a star's orbital
evolution is followed by simulating the random-walk of its velocity vector,
which approximates the effect of relaxation on the much shorter orbital
time-scale.  After each random-walk step the star is checked for entry into the
loss-cone. For further details see Freitag \& Benz (2002).
Comparison with $N$-body calculations have shown that in order to achieve
acceptable agreement between the two methods, the MC time-step must be
chosen rather small relative to the local relaxation time, with $dt \leq 0.01
T_{rel}(r)$ (Freitag et al. 2006). While choosing such a small time-step was
still feasible in the code of Freitag~\&~Benz~(2002), to enforce such a
criterion for all stars in our simulation would lead to a dramatic slow-down of
our code and notable spurious relaxation. The reason is that our code uses a
shared time step scheme, with the time-step chosen to be the smallest value of
all $dt_i=f~T_{rel}(r_i)$, where $f$ is some constant fraction and the subscript
$i$ refers to the individual star. In Freitag \& Benz (2002) each star is
evolved separately according to its local relaxation time, allowing for larger
time steps for stars farther out in the cluster where the relaxation times
are longer. In order to reduce the effect of spurious relaxation we are forced
to choose a larger $f$, typically around $f=0.1$. This has the consequence that
the time-step criterion is only strictly fulfilled for stars typically outside of $0.1 r_h$,
where $r_h$ is the half-mass radius of the cluster.
To arrive at the correct merger rate of stars with the IMBH, despite the
larger time step for the stars in the inner region, we apply the following
procedure: (i) for each star $i$ with $dt> 0.01 T_{rel}(r_i)$ we take $n=
dt/(0.01T_{rel})$ sub-steps. (ii) during each of these sub-steps we carry out
the random-walk procedure as in Freitag \& Benz (2002) (iii) after each sub-step
we calculate the star's angular-momentum $J$ according to the new velocity vector
(iv) we generate a new radial position according to the new J. 
By updating $J$ after each sub-step we approximately account for the star's
orbital diffusion in $J$ space during a full MC step, while neglecting
any changes in orbital energy. This is, at least for stars with low $J$,
legitimate (Shapiro \& Marchant 1978), while for the other stars the error might
not be significant as the orbital energy diffusion is still slower than the $J$
diffusion (Frank \& Rees 1976). A further assumption is that the cluster potential in
the inner cluster region does not change significantly during a full MC
step, which constrains the MC step size.

\section{Comparison to N-body Simulations}

\begin{figure} \begin{tabular}{cc}
  \includegraphics[width=0.38\textwidth,clip]{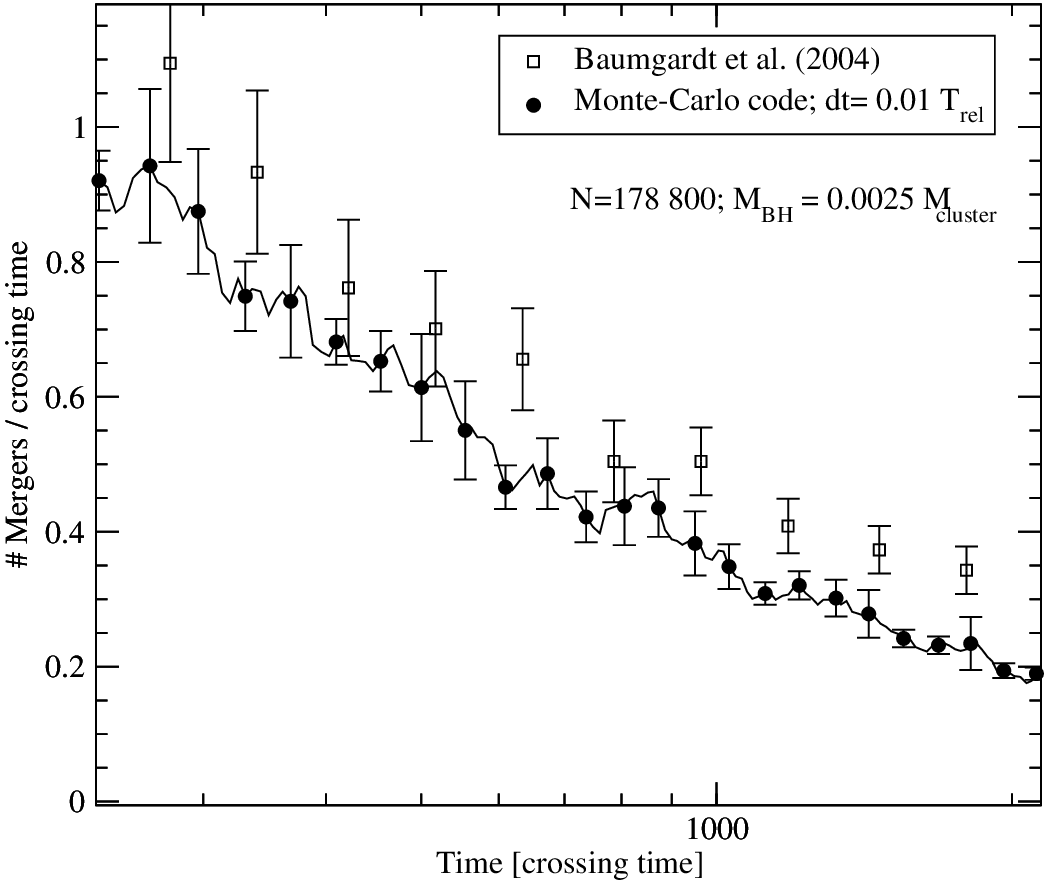}
  \includegraphics[width=0.345\textwidth,clip]{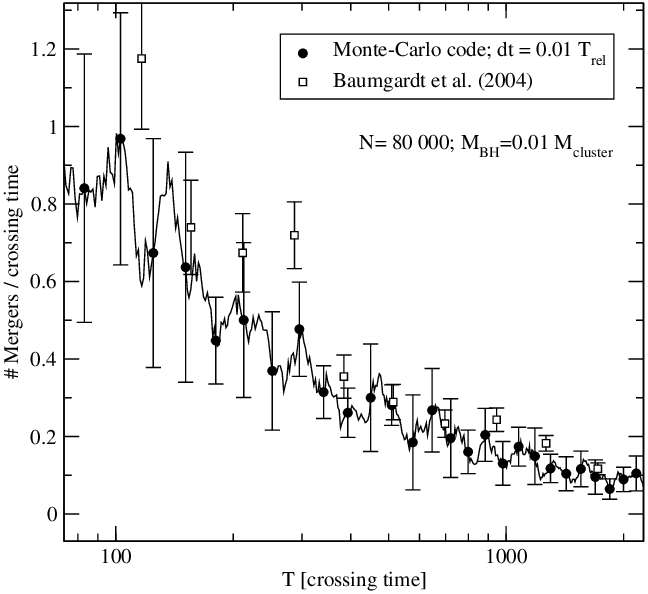} \end{tabular}
  \caption{Comparison of the stellar merger rate per crossing time with
  simulations of Baumgard et al. (2004). In both cases the evolution of
  single-mass clusters were calculated and the tidal radius of the IMBH was
  fixed to $1\times10^{-7}$ in N-body units. Full circles with error bars are
  results from our MC runs at selected times, while the solid line goes
  through all obtained points.} \label{mergerRates} 
  \end{figure}

In Fig. \ref{mergerRates} 
  the rates of stellar mergers with the central BH per crossing time from two
  of our single-mass cluster simulations are compared to the corresponding
  results of Baumgardt et al. (2004) (run16 and run2).  As can bee seen, the differences
  between our MC and the $N$-body results are within the respective error bars
  and thus in reasonable agreement with each other. However, the merger rates
  in the left panel of Fig.  \ref{mergerRates} seem to be consistently lower
  than in the $N$-body calculations. This might indicate that the agreement
  gets worse for other $M_{BH}/M_{Cl}$ than we considered here ($0.25\% -1\%$)
  and a different choice of time-step parameters for our MC code might be
  necessary in those cases. On the other hand, the differences might also be
  caused by differences in the initial relaxation phase before the cluster
  reaches an equilibrium state. This phase cannot be adequately modeled with a
  MC code because the code assumes dynamical equilibrium. Further comparisons to
  $N$-body simulations for different $M_{BH}/M_{Cl}$ and $N$ are necessary to
  test the validity of our method.  
  
  \section{Realistic Cluster Models} 
  
\begin{figure} \begin{tabular}{cc}
    \includegraphics[width=0.385\textwidth,clip]{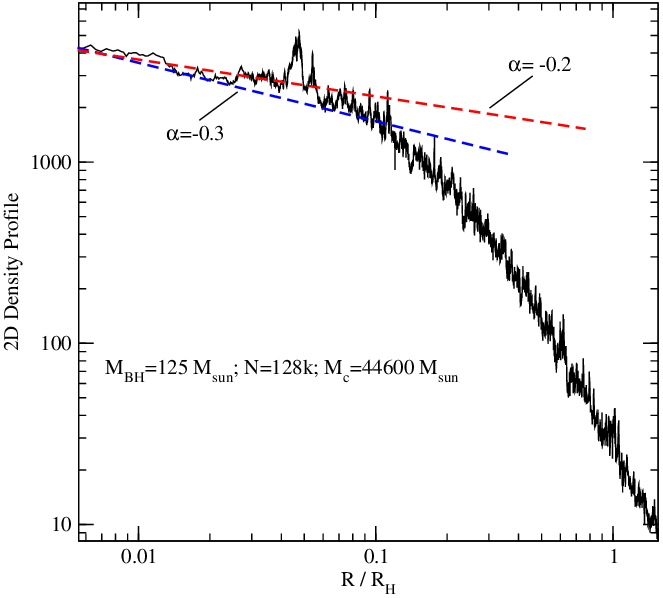}
    \includegraphics[width=0.385\textwidth,clip]{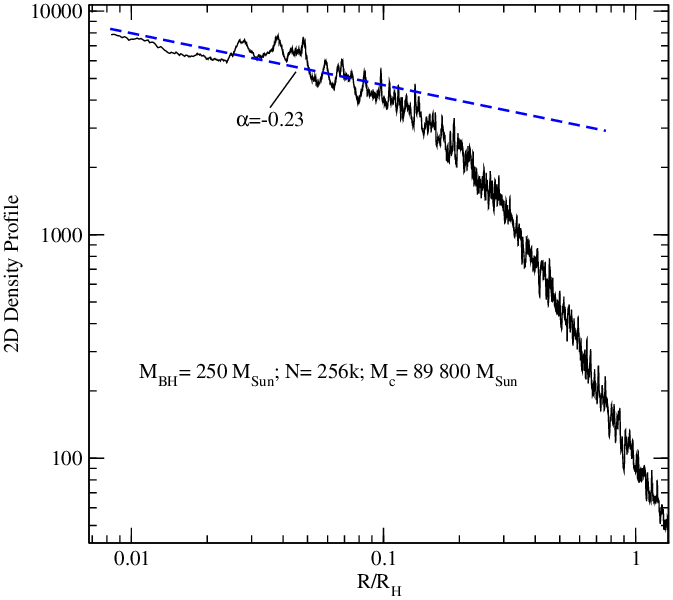}
  \end{tabular} \caption{Two-dimensional density profile of bright stars for two
  clusters with different numbers of stars and BH to cluster mass ratios. The
  dashed line in the right panel is a power-law fit to the inner region of the
  cluster, while the two dashed lines in the left panel are for orientation
  only.} \label{surfaceDensity} \end{figure}

  In order to compare our simulations to observed GCs additional physical
  processes need to be included. Here we consider two clusters containing
  $1.3\times10^5$ and $2.6\times10^5$ stars with a Kroupa-mass-function (Kroupa
  2001), and follow the evolution of the single stars with the code of
  Belczynski et al. (2002) (for all other parameter see Baumgardt et al. 2005).
  Fig.  \ref{surfaceDensity} 
  shows the two-dimensional density profiles of bright stars for the two
  clusters at an age of $12\rm{Gyr}$. The profile in the left panel can directly
  be compared to the corresponding result of Baumgardt et. al (2005) as $N$ is
  the same. As was expected from the discussion in \S \ref{ImprintsIMBH}, the
  profile shows only a very shallow cusp with a power-law slope $\alpha$ between
  $-0.2$ and $-0.3$, consistent with the $N$-body results. The right panel shows
  the resulting profile for a cluster that is similar but twice as massive
  and, consequently, has twice as many stars as in the $N$-body simulation.
  We obtain a very similar profile with $\alpha=-0.23$ which is very
  close to the average $\alpha=-0.25$ found in Baumgardt et al. (2005).
  Therefore, based on these very preliminary results, there seems to be no
  significant difference in cusp slopes for larger-$N$ clusters compared to
  small-$N$ ones, which means that no new candidate cluster 
  from the sample of Noyola \& Gebhardt (2006) can be identified in addition to
  those found by Baumgardt et al. (2005). However, the parameter space must
  be explored much further in order to confirm this finding.

  This work was  supported by NSF Grant AST-0607498 at Northwestern University.

  \newcommand{\apj}{\textit{ApJ}} \newcommand{\mnras}{\textit{MNRAS}}
  \newcommand{\aj}{\textit{AJ}} \newcommand{\aap}{\textit{A\&A}}
  \newcommand{\apjl}{\textit{ApJ} (Letters)}

\end{document}